\documentstyle[12pt]{article}

\begin{document}
\bibliographystyle{unsrt}

\begin{flushright} 
UMD-PP-96-59

January, 1996

hep-ph/9601203

\end{flushright}

\vspace{6mm}

\begin{center}

{\Large \bf $SU(5)\times SU(5)$ unification,
see-saw mechanism and R-conservation }\\ [6mm]

\vspace{6mm}

{\bf{R.N. Mohapatra\footnote{Work supported by the
 National Science Foundation Grant \#PHY-9119745 and a Distinguished
Faculty Research Award by the University of Maryland for the year 1995-96.}}}

{\it{ Department of Physics and Astronomy}}\\
{\it{University of Maryland}}\\

{\it{ College Park, MD 20742 }}

\end{center}

\vspace{4mm}

\begin{center}
{\bf Abstract}
\end{center}
\vspace{1mm}

   A supersymmetric grand unified model based on the gauge group
$SU(5)\times SU(5)$ is discussed . This model has the new feature that
the conventional see-saw mechanism for neutrino masses is embedded 
using the ${\bf (15,1)+(1,\overline{15})}$ representations.
This representation may have a better chance of arising from level two
compactification of superstring theories
 than the ${\bf \overline{126}}$-dimensional
representation used in the $SO(10)$ grand unified models. The model also
naturally suppresses all R-parity violating interactions.

\newpage

There is now a widespread belief that the
next step beyond the standard model may very well contain a
supersymmetric grand unified theory (SUSY GUT).  This speculation is
supported by the unification of gauge
couplings that has been observed to occur \cite{langa} at a scale around
$10^{16}$ GeV, with supersymmetry scale in the TeV region using
the precise values of the low energy couplings measured at
LEP and SLC.  Meanwhile, the superstring theories,
that have the potential to unify gravity with the strong and
electroweak forces, have been shown to 
lead to a variety of such SUSY GUT groups via an
appropriate compactification scheme. These facts have led to intensive
investigations of many grand unification scenarios, the simplest
and most notable ones
being the $SU(5)$\cite{nath} and $SO(10)$\cite{soten} schemes. 
The detailed aspects of the models of course are dictated by low energy
observations such as the spectrum of fermion masses, absence of baryon
and lepton number violating interactions etc. In our investigation, we will
focus on two properties, which we believe are desirable for any viable
SUSY GUT model: (i) there must be a simple way
to explain the smallness of neutrino masses; and (ii) R-parity violating
interactions that can lead to uncontrollable baryon and lepton number
violation must be absent or naturally suppressed. We would furthermore
require that the model contain Higgs representations that have a chance
of emerging from compactification of the heterotic string theory at
some Kac-Moody level.

The simplest way to accomodate massive neutrinos in GUT theories is via
the see-saw mechanism\cite{seesaw}, where the smallness of their
masses is linked to the largeness of the $B-L$ breaking scale. 
As far as automatic R-conservation is concerned, experience has shown that
it depends very sensitively on the nature of the gauge
symmetry and the Higgs content of the
model\cite{moh}. Of course one may take the point of view that the process of
string compactification may yield R-parity as a discrete symmetry. However,
this property is much harder to demonstrate in a string theory.
Furthermore, if R-parity arises
as global symmetry, one still has to worry about possible Planck suppressed
R-violating interactions of gravitational origin which can lead
to undesirable and often catastrophic effects. It may therefore be preferable
to rely on the gauge group and Higgs representations as a way to guarantee
R-parity as an automatic symmetry of the theory.

Let us now come to specific models. In the simplest models based on the
$SU(5) $ group, $B-L$ is not a gauged symmetry, and therefore, they are
not suitable for understanding non-zero neutrino masses. Furthermore,
the $SU(5)$ theory also allows arbitrary strengths for R-violating
interactions.  This takes us to the next class of models
 based on the $SO(10)$ group, where (i)
the see-saw mechanism can be implemented by using the $\bf{( 126
+\overline{126})}$ representations to break the $B-L$ symmetry, and 
(ii) also as noted in \cite{lee}, this model has the property that it
leads to automatic suppression of all R-violating
interactions, which satisfies our second requirement above. However,
it appears increasingly unlikely
\cite{lykken} that the $\bf{(126+\overline{126})}$ representations
can emerge from the
compactification of heterotic string models. \footnote{Strictly the authors of
Ref.\cite{lykken} have proved this for fermionic compactification schemes.}
Therefore such scenarios will be disfavored if one believes in
superstring theories as the final theory of nature. The simplicity
of the conventional seesaw mechanism and the requirement of automatic
R-conservation are so appealing as phenomenological requirements that
it is useful to seek alternative SUSYGUT frameworks which not only
have both these properties but which also have a better chance of emerging
from superstring models .

 In this letter, we present a SUSY GUT model
based on the gauge group $SU(5)\times SU(5)$\footnote{A different class
of $SU(5) \times SU(5)$ string-embeddable GUT model has recently been
proposed in Ref.\cite{barb}; our model is very different from this model
not only interms of the fermion content but also symmetry breaking as well
as of course in its implementation of the see-saw mechanism.}
 which  provides an alternative way to implement
the see-saw mechanism using the representations
${\bf ( \overline{15}, 1)+(1, 15)}$. This may be more amenable to superstring
embedding because, it has now been shown that for a single $SU(5)$ group
in a level two string compactification, 
there appear ${\bf 15}$-dimensional representations
\cite{lykken} and it is quite likely that for the $SU(5)\times SU(5)$
case, the representations we use will appear. So far only level one 
$SU(5)\times SU(5)$ models have been studied\cite{finnelli} and perhaps
the considerations of the present paper will motivate a study of these models
at level two. At low energies, this model coincides with the usual minimal
supersymmetric standard model ( MSSM ) .
 Another property of our $SU(5)\times SU(5)$ model is that
the R-parity violating interactions are naturally suppressed. 

\vspace{4mm}
\noindent{\bf The $SU(5)\times SU(5)$ model:}
\vspace{2mm}

We assume the gauge group to be $SU(5)_{A}\times SU(5)_{B}$ with the
associated gauge couplings denoted by $g_A$ and $g_B$ respectively. We
will se later that phenomenologically reasonable unification scale 
consistent with the low  energy precision measurement of the standard
model gauge couplings will require that at the GUT scale the two couplings are
unequal. Such a scenario requires that the discrete symmetry that transforms 
one $SU(5)$ group to the other is broken at string scale.

We will assign the matter superfields to transform as 
${\bf (\overline{5}+~10,~1)+~(1,~5+\overline{10})}$ for 
each generation. This implies that we must have
extra fermions beyond those present in the standard model. We denote them by
$( U,~U^c,~D,~D^c,~E,~E^c)$; of these the $(U,U^c,D,D^c)$ are the heavy
vector like analogs of the familiar up and down quarks respectively ( and 
therefore have obvious $SU(3)_c$ color transformation properties) whereas
the $E,E^c$ are color singlet singly charged heavy fermions. These extra
fermions also come three varieties corresponding to the three generations
of known quarks and leptons. Note that there is no heavy vectorlike analog
of the neutrinos. The assignment of these fermions to the representations of
the gauge group are given below:

\begin{eqnarray}
\psi~=\left(\begin{array}{c}
D^c_1\\
D^c_2\\
D^c_3\\
e^-\\
\nu \end{array}\right);~
\chi~=\left(\begin{array}{ccccc}
0 & U^c_3 & -U^c_2 & u_1 & d_1 \\
-U^c_3 & 0 & U^c_1 & u_2 & d_2\\
U^c_2 & -U^c_1 & 0 & u_3 & d_3 \\
-u_1 & -u_2 & -u_3 & 0 & E^+\\
-d_1 & -d_2 & -d_3 & -E^+ & 0\end{array} \right);~
\end{eqnarray}

Similarly, the fermions in ${\bf (1, {5}+\overline{10})}$ denoted by $\psi^c$
and $\chi^c$ can be written as :
\begin{eqnarray}
\psi^c~=\left(\begin{array}{c}
D_1\\
D_2\\
D_3\\
e^+\\
\nu^c\end{array}\right);~
\chi^c~=\left(\begin{array}{ccccc}
0 & U_3 & -U_2 & u^c_1 & d^c_1\\
-U_3 & 0 & U_1 & u^c_2 & d^c_2 \\
U_2 & -U_1 & 0 & u^c_3 & d^c_3 \\
-u^c_1 & -u^c_2 & u^c_3 & 0 & E^-\\
-d^c_1 & -d^c_2 & - d^c_3 & -E^- & 0\end{array}\right).
\end{eqnarray}

Note that below the scale where the heavy vectorlike fermions ( i.e,
$U,~D,~E$) become massive, the fermion content is same as in the left-right 
symmetric models. To make this point transparent, we discuss the symmetry 
breaking of the GUT group to the standard model gauge group below.
We assume the Higgs fields of the model to transform as follows: there
are two sets of multiplets belonging to 
${\bf ( 5,~\overline{5})+(\overline{5},~5)}$
representations ( denoted by $H_{1,2}+\overline{H}_{1,2}$ ); 
one set belonging to ${\bf (24,~1)+(1,~24)}$
 ( denoted by $\Phi_A+\Phi_B$ ) and another set transforming as 
${\bf ( 15,~1)+ (1,~\overline{15})}$ ( denoted by $S_A~+S_B$ )
and ${\bf (\overline{15}~,1)+(1,~15)}$ (denoted by $\overline{S}_A+
\overline{S}_B$ ). The first point to note is
that all these Higgs representations have good chance of arising from
level two fermionic compactification of the heterotic string theory as already 
mentioned earlier since both the representations
 {\bf 15} and {\bf 24} have already been shown to appear in level two
$SU(5)$ string GUT models.

Turning to symmetry breaking and fermion masses, we assume that
$H_1$ has GUT scale vev with the pattern $ \langle H_1 \rangle = diag
( V,V,V,0,0)= \langle \overline{H}_1\rangle$ so that the gauge group  
breaks down to  $SU(3)_c\times SU(2)_L\times SU(2)_R \times U(1)_{B-L}$. 
At the next stage, we assume that
$S_B$ has a vev along the $\nu^c\nu^c$ direction i.e. $\langle S_{B,55}\rangle
= V_R$ so that the gauge group below $V_R$ is the standard model group.
As we will show the see-saw mechanism for neutrino masses arises at this
stage.
The final breaking of the standard model is achieved by giving vev to
the $H_2$ as $\langle H_2 \rangle = diag(0,0,0, \kappa, \kappa')$.
We also assume that $\overline{H}_2$ has a similar vev pattern.
In principle, the seesaw scale $V_R$ could be lower than the GUT scale 
Here for simplicity, we will focus on a scenario with $V=V_R$.

Having provided the general outline and the symmetry breaking of the model,
 let us now discuss the decoupling of the heavy fermions and the generation
of the various light fermion masses in the theory. The two crucial vev's
for this discussion are those of $H_{1,2}$ given above. The heavy fermion
( i.e. $U,D,E$ ) masses arise from the following couplings in the
 superpotential: $\psi H_1 \psi^c$ and $\chi \overline{H}^2_1 \chi^c/{M}$
where $M$ could either be a scale corresponding to new physics between
$M_{GUT}$ and the Planck scale or the Planck scale itself i.e.
$M=M_{Pl}/{\sqrt{8\pi}}~\simeq 10^{18}~GeV$. It
is easy to check that these two couplings give masses of order $M_{GUT}$
and $M^2_{GUT}/M$ to the $D$  and $U$ colored fermions respectively. If
we choose $M_{GUT}\simeq 10^{16}$ GeV, we could choose $M\simeq 10^{17}$
GeV or so so that $M_U\simeq 10^{15}$ GeV.
The mass of the heavy lepton $E$ arises from the non-renormalizable 
coupling $\varepsilon_{ijklm}\chi^{ij}H^k_{1,p}H^l_{1,q}H^m_{1,r}\chi^{c}_
{st}\varepsilon^{pqrst}/ {M^2}$. We get $M_E\simeq 10^{14}$ GeV. Thus ,
the vector like fermions decouple from the low energy spectrum.

Let us now turn to the light fermion masses. Since they arise from
the breaking of the standard model gauge group, they will involve
the vev of $H_2$ given above. The relevant terms in the superpotential
that give rise to these masses  are  $h_e \psi H_2 \psi^c$, 
$h_q \chi \overline{H}_2 \overline{H}_1 \chi^c / M $ 
where $h_{e,q}$ are $3\times 3$
matrices in the generation space; the $h_e$ term gives rise to the
charged lepton masses and the Dirac  masses for the neutrinos and the
$h_q$ term gives rise to the mass matrices for the up and the down quark
masses. Note that since the up and down quark masses at this stage are
proportional to each other, the CKM mixing angles vanish. We also note that
the $33$ element of $h_q$ must be of order 3 to 10 in order to understand
why the top quark is so heavy.  In order to generate quark mixing angles,
we include next higher terms in the superpotential of the form 
$\chi \overline{H}_1 \overline{H}_2 \chi^c H_1 \overline {H}_1 / {M^3}$. 
The coupling 
matrix in front of the above term leads to a misalignment between the
up and down sectors leading to non-vanishing CKM angles. Moreover,
this being a higher order term in $M_{GUT}/M$, also naturally explains
why the mixing angles between the quarks must be small.

An important point to note here is that in our model there is no mixing
between the superheavy vector like quarks and leptons and the known light
fermions. In this respect our model differs from several recent 
works\cite{cho} , who had very similar fermion assignment to ours but
had the heavy-light fermion mixing as an essential ingredient.

Let us now discuss the implementation of the see-saw mechanism in our model.
Crucial for this purpose is the symmetric ${\bf ( 15, 1)+(1, \overline{15})}$
representations. In our superpotential we include their coupling to the 
fermions given by $f(\psi\psi S_A+\psi^c \psi^c S_B)$. As already mentioned,
the vev $\langle S_{B,55}\rangle= V_R$ breaks the $B-L$ symmetry and
in that process gives  Majorana masses to the right-handed neutrinos
of magnitude $f V_R$. Since we already showed that neutrinos have Dirac
masses from their $\psi H_2\psi^c$ coupling, we now have all the ingredients
of the conventional see-saw mechanism\cite{seesaw} leading to the
usual see-saw formula for neutrino masses i.e. $m_{\nu_i}\simeq m^2_{\nu^D}
/f_i V_R$. The gauge symmetry allows a nonrenormalizable term of the
form $S_A H^2_2 S_B/M$, which induces the vev for $S_{A,55}$ of order
$\approx \kappa^2/M$, which is of order $10^{-5}$ eV and hence too small
to effect the usual see-saw formula. The symmetric {\bf 15}-
dimensional multiplet
plays exactly the same role as the ${\bf \overline{126}}$ multiplet in the
$SO(10)$ models except that it has a chance to arise from level two 
compactification of heterotic string models unlike the ${\bf \overline{126}}$.
 
\vspace{6mm}
\noindent{\bf Gauge coupling unification:}

\vspace{3mm}

Let us now turn to gauge coupling unification in the model and the mass scales
for the symmetry breaking at various stages. It was noted in \cite{cho}, that
if we assume the gauge couplings for the two $SU(5)$ groups to be equal
( i.e. require an exact discrete symmetry that transforms $SU(5)_A$ to
$SU(5)_B$ ), then the model predicts very small value for $sin^2\theta_W
= 3/16$ at the GUT scale and since $sin^2\theta_W$ 
decreases in general at lower
scales, such a scenario is in gross disagreement with observations. On
the other hand in string inspired models it is conceivable that the 
discrete symmetry that guarantees that the two $SU(5)$ couplings
are equal, is broken below the string scale. This could for instance
happen if there are singlet fields odd under the above discrete symmetry,
in which case, non-renormalizable couplings involving this field and the
two SU(5) gauge fields could also lead to splitting between the two gauge
couplings. We will therefore work with a scenario where this happens
(i.e. where $g_A\neq g_B$ at the GUT scale). If
we denote the ratio of the two fine structure constants 
for the two $SU(5)$ groups by $y\equiv({{\alpha_A}/{\alpha_B}})$, 
then we get $ sin^2\theta_W(M_{GUT})={{3}\over{8(1+y)}}$; 
if we then chose $y\ll 1$, then the GUT scale value for $sin^2\theta_W$
approaches that of the $SU(5)$ or $SO(10)$ prediction for it and
one can obtain a value for it at the scale $M_Z$ in agreement with 
observations. Below we present an example of such a scenario. 

The equations relating the gauge couplings at $M_Z$ to those at the
GUT scale are given by:

\begin{eqnarray}
\alpha^{-1}(M_Z)~=~{{5}\over{13}}\alpha^{-1}_A~+~{{8}\over{13}}\alpha^{-1}_B
+~b_1 U +~c_1 R \\
\alpha^{-1}_2(M_Z)~=~\alpha^{-1}_A + b_2 U + (b'_2-b_2)~R\\
\alpha^{-1}_3(M_Z)~=~{{1}\over{2}}(\alpha^{-1}_A+\alpha^{-1}_B)+ b_3 U
+ (b'_3-b_3) R\\
\end{eqnarray}   

In the above equations, we have denoted $U={{1}\over{2\pi}}ln(M_{GUT}/M_Z)$
and $R={{1}\over{2\pi}}ln(M_{GUT}/M_R)$ , with $M_R$ denoting the scale at
which the symmetry $SU(2)_R\times U(1)_{B-L}$ breaks down to $U(1)_Y$ and
$M_{GUT}$ as already noted stands for the unification scale. The coefficients
in front of $U$ and $R$ are model dependent and represent the way the
gauge couplings evolve in different models. We have assumed only a single
intermediate scale below $M_{GUT}$ corresponding to the symmetry 
$SU(3)_c\times SU(2)_L \times SU(2)_R \times U(1)_{B-L}$ 
(denoted by $G_{3221}$ ) above $M_R$
and also that the $SU(2)_L$ group is embedded only in $SU(5)_A$ group.
In this case we have: $b_1={{3}\over{13}}\Sigma\left({{Y}\over{2}}\right)^2$,
$c_1={{3}\over{13}}b_{2R}+{{10}\over{13}}b_{BL}-b_1$; $b_{2L}$ and $b'_{2L}$
are $SU(2)_L$ beta function coefficients above and below the scale $M_R$ and
$b_3$ and $b'_3$ are the $SU(3)_c$ beta function coefficients above and
below $M_R$ ( their values are half the usual $SU(3)_c$ coefficients
due to the fact that $SU(3)_c$ arises as the diagonal sum of the two
$SU(3)$'s in the $SU(5)_{A,B}$)
 and $b_{BL}=\left({{3}\over{10}}\right)\Sigma \left({{B-L}\over
{2}}\right)^2$.

As an example of a scenario , we assume $M_{GUT}=M_R$ and MSSM
multiplet content below $M_{GUT}$ except that we require the
Higgs fields in the representation $\bf {(1, 3 ,\pm 2)}$ 
and one color octet ${\bf (8,1,0)}$ under the standard model group
$SU(3)_c\times SU(2)_L\times U(1)_Y$ to have intermediate to low mass.
Such a choice is completely compatible with low energy phenomenology.
For this case, we find $M_{GUT}\approx 10^{16}$ GeV in the one loop
approximation with $\alpha^{-1}_A\simeq 17$ and $\alpha^{-1}_B\simeq 1$
the triplet and octet fieds with same mass $\approx 10^{10.5}$ GeV.
Another scenario which also yields an $M_{GUT}=M_R\simeq 10^{16}$ GeV
is one with only a single $B-L$ neutral $SU(2)_L$ triplet with mass around
$10^{11}$ GeV giving $\alpha^{-1}_A=20.7$ and $\alpha^{-1}_B=2.4$.
The latter case is preferable from one loop perspective since both gauge
couplings are in the perturbative region.
 We emphasize that the spectra chosen in both these examples arise
 from the particle content of the theory
by appropriate fine tuning. We expect that once two loop and the 
threshold corrections are included, the GUT scale could easily reach
the range used in the fermion mass discussion of the paper.

It may be worth pointing out at this stage that, the light doublets
needed for electroweak  symmetry breaking arise in this model from the
combination of terms $H_2\Phi\overline{H_2}+\mu H_2\overline{H_2}$ by
appropriate fine tuning of the parameter $\mu$. This would leave us 
with two pairs of standard model doublets; one of the pairs becomes
very heavy due to the nonrenormalizable interaction $\overline{H}^i_{1,p}
\overline{H}^j_{1,q}\overline{H}^l_{1,r}\overline{H}^m_{2,s}\overline{H}^n
_{2,t}\varepsilon^{pqrst}\varepsilon_{ijlmn}/{M^2}$.

\vspace{4mm}

\noindent{\bf R-parity breaking, proton decay etc:}

\vspace{4mm}

In the present model, R-parity is automatically conserved even in the
presence of nonrenormalizable Planck scale suppressed terms. To see this,
recall that in the conventional $SU(5)$ SUSYGUT model, one source of
R-parity breaking interactions is the term in the superpotential involving
$10~\overline{5}~\overline{5}$ terms ( all fields are matter fields ) which
leads to terms of type $u^cd^cd^c$ and $QLd^c$ . However in our case
due to the assignment of heavy fermions, the analogous terms give rise
to operators of type $QLD^c$, $U^cD^cD^c$ which do not lead to
R-parity violation involving light fermions. Furthermore, there is no
mixing between the heavy and light quarks in this model due to the existence
of an exact global symmetry $(-1)^{B+H-L}$ where all quarks ( both light
and heavy ) have the usual baryon number $B=1/3$; the quantum number
$H$ is +1 for heavy quarks and leptons and zero for light fermions.
Thus R-parity is exactly conserved and as a rseult of this , the
lightest supersymmetric particle (the LSP ) is absolutely stable and
can play the role of dark matter of the universe. 

Coming to proton decay, again because there is no mixing between
the heavy and light quarks, there are no leading order contributions
to proton decay until we include
non-renormalizable Planck scale induced terms. In the
lowest order in $M^{-1}_{Pl}$, proton decay arises from the following operator:
$\chi\chi\chi \psi $  which gives an operator of type $QQQL$ with
coefficient $\lambda/{M_{Pl}}$.
After gluino and wino dressing, it would lead to the
four-fermion proton decay operator with strength $(\lambda g^2 M_{gaugino})/
(16\pi^2 M^2_{sq} M_{Pl})$. For $\lambda\simeq 10^{-5}$ and other
parameters being reasonable, this leads to
proton lifetime long enough to be consistent with observations.
Due to the presence of the unknown coupling parameters, it is not
possible to make a more definitive statement about the proton lifetime.

In conclusion, we have pointed out that a SUSY GUT theory based on the
$SU(5)\times SU(5)$ group has two properties highly desirable of
a GUT model: (i) it can embed the see-saw mechanism 
for neutrino masses using multiplets that
have a chance of emerging from the superstring theories; and(ii) the
model has automatic R-parity conservation thus guaranteeing that the
LSP ( which is supposed to play the role of dark matter in the universe )
is indeed truly stable without any extra theoretical assumptions.
We have also analysed the coupling constant unification in this class
of theories in the one-loop approximation and showed that realistic
scales can emerge provided the two $SU(5)$ couplings are different from
each other at the GUT scale.

\end{document}